\documentstyle[sprocl,amssymb,amsmath,calc]{article}

\def\be{\begin{equation}}
\def\ee{\end{equation}}
\def\bea{\begin{eqnarray}}
\def\eea{\end{eqnarray}}

\begin{document}
\title{A FRESH LOOK AT THE ADHESION APPROXIMATION}

\author{T. BUCHERT}

\address{Theoretische Physik, Ludwig-Maximilians-Universit\"at\\
Theresienstr. 37, D-80333 M\"unchen, Germany}


\maketitle\abstracts{
I report on a systematic derivation of the phenomenological 
``adhesion approximation'' from gravitational instability
together with a brief evaluation of the related status of
analytical modeling of large--scale structure.
}

\vspace{-0.3cm}
\noindent
Hectored by steadily improving computer power, N--body simulations
suggestively establish themselves as the final answer to the modeling of
large--scale structure in the Universe, admitting their limitations only in
covered dynamical range and lack of ``input physics''. Still,
the analytical approach provides for a complementary way of 
understanding. Ironically, ``understanding'' large--scale structure formation
is itself becoming a discipline of N--body computing. 
Notwithstanding, a brief reminder of the line of 
development of analytical modeling is in order.

Well before we finally saw the structures, which are generically encountered
around us, also in the Universe, Zel'dovich put forward a kinematical
picture of large--scale structure formation into that what we nowadays call
network, web, filaments, sheets and voids.
Here, understanding preceeded observation and finalized the necessity of
departure of
N--body simulations from spherical symmetry. Thereafter, almost the 
same picture is repeatedly presented.
We learnt that N--body results down to the scales where
we may consider them reliable can be understood, if not mirrored, by
refining Zel'dovich's step (e.g. Lagrangian perturbation theory,
now conquering the general--relativistic regime; optimized,
i.e., high--frequency truncated Lagrangian schemes). 
One may consider it good or bad news that N--body results 
in gross terms agree with such simple models 
(for further reading see the review by Sahni \& Coles$^1$ and 
references therein). 

Analytical insight has concentrated on the issue of ``shell--crossing'' or 
``multi--streaming'', which is not a problem of N--body computing, but of
analytical modeling using ``dust matter''.
Known for a long time outside of cosmology, Shandarin has particularly 
pushed the idea that this shortcoming after caustic formation can be 
overcome by Burgers--type equations for an ``adhesive'' modeling of 
structures$^2$: an additional force proportional to the Laplacian of
the peculiar--velocity is capable of stabilizing collapsed structures. 

In a recent work$^3$ we have taken up this line
by demonstrating that a systematic
derivation of the adhesion model from gravitational instability is feasible. 
It is important to realize that also collisionless matter features 
pressure--like forces arising from the action of multi--streaming, a 
well--known fact in stellar systems theory. Forming velocity moments of the
collisionless Boltzmann equation, we arrive at the following
hierarchy of evolution equations:
\vspace{-0.1cm}
\begin{gather}
\partial_{t} \varrho + (\varrho {\bar v}_{i})_{,i} = 0 \;\;\; ,\\
\partial_{t} {\bar v}_{i} + {\bar v}_{j} {\bar v}_{i,j} = g_{i} - 
{1 \over \varrho} \Pi_{ij,j} \;\;\; ,\\
\partial_{t} \Pi_{ij} + {\bar v}_k \Pi_{ij,k} + {\bar v}_{k,k} \Pi_{ij} = 
- \Pi_{ik} {\bar v}_{j,k} - \Pi_{jk} {\bar v}_{i,k} - L_{ijk,k} \;\;\;,
\end{gather}

\noindent
and so on, where $\Pi_{ij} = \varrho \langle (v_{i}-{\bar v}_{i})(v_{j}-{\bar
v}_{j}) \rangle$ is the tensor of velocity dispersion taking the multi--stream
force into account, ${\bar v}_i$ and
$g_i$ are the components of the mean
velocity and the gravitational field strength obeying Newton's field 
equations, respectively, and $\varrho$ is the mass density. 
We have suggested to close this hierarchy by assuming that velocity 
dispersion is small, but non--vanishing. Hence, beyond Jeans' equation (2) 
we only keep the evolution 
equation for the dispersion tensor (3), but neglect the contribution from the
third--rank tensor $L_{ijk,k}$.
Two further crude$^4$ approximations then lead us to the key equation of the 
adhesion approximation: we only consider the isotropic part of the dispersion
tensor, $\Pi_{ij} \approx p \delta_{ij}$ and, after transforming  
to peculiar--fields, we solve the resulting equation 
by iteration and find the first iterate to obey$^3$:
\vspace{-0.1cm}
\begin{equation}
\frac{d \tilde{\bf u}}{d b} = {\mu} \Delta_{\bf q} \tilde{\bf u} \;\;\; ; 
\;\;\;  \frac{d}{d b} := \frac{\partial}{\partial b} + 
\tilde {\bf u} \cdot \nabla_{\bf q} \;\;\;,\;\;\;\tilde{\bf u}: = 
\frac{{\bf u}}{a\dot{b}}\;\;\;,\;\;\;{\bf q}:= \frac{{\bf x}}{a}\;\;\;, 
\end{equation}

\noindent
where $\bf u$ is the peculiar--velocity, $a(t)$ is the scale factor of the
background cosmology, $b(t)$ is the growing mode solution 
of the linear theory, and $\mu$ is
a function of time and density. This density dependence disappears for 
$p \propto \varrho^2$, whereas from (3) $p \propto \varrho^{5/3}$.
In view of the restrictions done there is plenty of room for a generalization of
the adhesion approximation, which may lead us to a deeper understanding
of structure formation in the nonlinear regime.

\section*{Acknowledgments}
\vspace{-0.2cm}
This work is supported by the {\em Sonderforschungsbereich SFB 375
f\"ur Astro--Teilchenphysik der Deutschen Forschungsgemeinschaft}. 
\vspace{-0.1cm}
\section*{References}

\noindent
1. V. Sahni, P. Coles (1995) Physics Reports 262, 1.

\noindent
2. S.N. Gurbatov, A.I. Saichev, S.F. Shandarin, (1989) M.N.R.A.S.
236, 385.

\noindent
3. T. Buchert, A. Dom\'\i nguez, A.\& A., submitted (1997).

\noindent
4. T. Buchert, A. Dom\'\i nguez, J. P\'erez--Mercader, Ap.J., submitted (1997).

\end{document}